\documentclass[conference]{IEEEtran}
\IEEEoverridecommandlockouts
\usepackage{cite}
\usepackage{amsmath,amssymb,amsfonts}
\usepackage{algorithmic}
\usepackage{graphicx}
\usepackage{textcomp}
\usepackage{xcolor}
\def\BibTeX{{\rm B\kern-.05em{\sc i\kern-.025em b}\kern-.08em
    T\kern-.1667em\lower.7ex\hbox{E}\kern-.125emX}}
\begin{document}

\title{New binary quantum codes from Hermitian dual-containing Quasi-cyclic codes\\

\thanks{This research is funded by the National Natural Science Foundation of China under Grant No.U21A20428, Natural Science Foundation of
Shaanxi under Grant No.2025-JC-YBQN-070. }
}

\author{\IEEEauthorblockN{1\textsuperscript{st} Liangdong Lu\textsuperscript{*}}
\IEEEauthorblockA{\textit{Department of Basic Science } \\
\textit{Air Force Engineering University}\\
Xi'an, Shaanxi,  P. R. China \\
Kelinglv@163.com}
\and
\IEEEauthorblockN{1\textsuperscript{st} Gaochi Zhang}
\IEEEauthorblockA{\textit{Information and navigation college} \\
\textit{Air Force Engineering University}\\
Xi'an, Shaanxi,  P. R. China \\
1993313930@qq.com}
\and
\IEEEauthorblockN{2\textsuperscript{nd} Ganyu Feng}
\IEEEauthorblockA{\textit{Information and navigation college} \\
\textit{Air Force Engineering University}\\
Xi'an, Shaanxi,  P. R. China \\
1982150416@qq.com}
\and
\IEEEauthorblockN{3\textsuperscript{rd} Wenzheng Ma}
\IEEEauthorblockA{\textit{Information and navigation college} \\
\textit{Air Force Engineering University}\\
Xi'an, Shaanxi,  P. R. China \\
280606050@qq.com}

}

\maketitle

\begin{abstract}
In this paper, we introduce a class of dual-containing 2-generator quasi-cyclic codes over $GF(4)$.
	By Hermitian construction, we show computation constructions of binary  quantum codes with good parameters 
from quaternary quasi-cyclic codes.   Furthermore,   3 new quantum codes $[[42,10,9]]$, $[[74,,36,9]]$ and 
$[[78,40,9]]$ are  derived from Hermitian dual-containing
 quasi-cyclic codes by  an exhaustive search, 
along with 24 newly derived codes that enhance the previously best-known lower bounds on minimum distance.
\end{abstract}


\section{Introduction}
Quantum error-correcting codes (QECCs), which safeguard delicate qubits against noise, are crucial for enabling quantum computing and quantum communication.
 The theory of QECCs has had a significant development since the initial works of Shor \cite{shor1995scheme} 
 and Steane \cite{Steane1996}.	 Ref. \cite{Calderbank1998} establishes a fundamental link between quantum error-correcting codes and classical quaternary codes..
Later, In Refs. \cite{Rains1999,Ashikhmin2001,Ketkar2006}, there is  
a correspondence to self-orthogonal additive codes  over  ${\mathbf{F}}_{q^{2}}$ . 
In recent years, Calderbank-Shor-Steane (CSS) construction and Hermitian 
construction are the  famous constructions of quantum codes. 
Many excellent quantum codes are constructed from self-orthogonal codes (also known as dual-containing codes) over finite fields. These self-orthogonal codes can be efficiently obtained using algebraic codes, 
including cyclic codes, constacyclic codes, quasi-cyclic codes, and algebraic geometry (AG) codes, and so on \cite{Kai2014ConstacyclicCA,Li2019NewQC,Guardia2012OnTC,feng2004finite}.

Quasi-cyclic code has a rich algebraic structure and excellent properties, 
which is a natural extension of cyclic code.\cite{daskalov2003new,Sguin2004ACO,Akre2021NewBA,Ling2005OnTA,Barbier}. 
If $d$ is the greatest known value for which an $[n,k,d]$ code exists, then the code $\mathcal{C}$$=[n,k,d]$ is referred to as a {\it best-known } code.
 Since many new classical {\it best-known }  codes  are constructed from quasi-cyclic codes, 
much works has been focused on constructing quasi-cyclic codes
\cite{daskalov2003new,siap2000new,chen2018some}. 
Due to these excellent properties, scholars have considered whether quasi-cyclic codes can be applied to the 
construction of quantum codes.  Hagiwara et al.\cite{hagiwara2011spatially} proposed  
quantum LDPC codes from quasi-cyclic codes in 2011.
The original method for constructing quantum codes from 
quasi-cyclic codes about symplectic, Euclidean  
and Hermitian inner products is presented in \cite{galindo2018quasi}.

 In  \cite{ezerman2019good}, Ezerman et al. use the quantum construction X to construct quantum codes by 
  quasi-cyclic codes
with large Hermitian hulls.
In Refs. \cite{lv2019new,lv2020explicit}, Lv et al. obtained some new
quantum codes from dual-containing one(or two)-generator quasi-cyclic codes
via the symplectic construction and the Hermitian construction
in turn. One of  difficult problems is that  these codes needed to calculate the exact dual distances
of these quasi-cyclic codes.
In  \cite{Guan2022QC}, Guan et al. provided the dimensionality of 2-generator quasi-cyclic codes
 under some specific conditions, but they did not prove 
 the dimensionality of 2-generator quasi-cyclic codes under general conditions.

Building on the aforementioned studies, we introduce a novel approach for creating quantum codes using 2-generator self-orthogonal (or dual-containing) quasi-cyclic codes of index 2. The quantum codes developed in this work 
exhibit improved parameters compared to those currently documented in  \cite{Grassl:codetables}.	
Here is a brief overview of the structure of this paper.  
Section 2 covers the basics of quasi-cyclic codes and quantum codes.  
In Section 3, a category of 2-generator quasi-cyclic codes is introduced along with a sufficient condition for self-orthogonality with respect to the Hermitian inner product.  
By applying the Hermitian construction, numerous new binary quantum codes are developed.

\section{Preliminaries}

In this section, we present some fundamental concepts related to quaternary linear codes, quasi-cyclic codes, and quantum codes.
Denoted the
Galois field with four elements $F_{4}=\{0,1,\omega,\varpi\}$. 
$\varpi=1+\omega=\omega^{2},\omega^{3}=1$, and the conjugation is
defined by $\bar x=x^{2}$ for $x\in F_{4}$. A classical linear code $\mathcal{C}$ of length $n$ over the field $F_{4}$ is an non-empty subset of  $F_{4}^{n}$, and is denoted by $[n,k,d]_{4}$.
Given two vectors $\vec{u}=(u_{0}, \ldots, u_{n-1})$ and
$\vec{v}=(v_{0},\ldots, v_{n-1})$ are vectors in $ F_{4}^{n}$, the Hermitian inner product between them is defined as
$\langle\vec{u}, \vec{v}\rangle_{h}=\sum_{i=0}^{n-1}  u_{i} v^2_{i} $.
The weight of $\vec{u}$ is the number of nonzero coordinates in $\vec{u}$, which is denoted by $wt(\vec{u})$.
The minimum non-zero Hamming weight of $\mathcal{C}$ is $d(\mathcal{C})=\min \left\{wt(\vec{u}) \mid \vec{u} \in \mathcal{C}, \vec{u} \neq 0\right\}.$
For linear codes, the minimum distance $ d $ is the smallest non-zero Hamming weight among all codewords.
Let $\mathcal{C}^{\perp_{h}}=\{\vec{v} \in F_{4}^{n} \mid\langle\vec{u}, \vec{v}\rangle_{h}=0, \forall \vec{u} \in \mathcal{C}\}$ be the Hermitian dual code of $\mathcal{C}$. 
If $\mathcal{C} \subset C^{\perp_{h}}$, then we can say $\mathcal{C}$ is a Hermitian self-orthogonal code and $\mathcal{C}^{\perp_{h}}$ is a Hermitian dual-containing code.

\subsection{Quasi-cyclic code }
 We define the quotient ring  $\mathcal{R}=F_{2}[x] /\left\langle x^{n}-1\right\rangle$. 
 If $\mathcal{C}$ is generated by a monic divisor 
 $g(x)$  of  $x^{n}-1$, i.e., $\mathcal{C}=\langle g(x)\rangle$ and $g(x)\mid x^n-1$, 
 then $g(x)$ is called generator polynomial of $\mathcal{C}$.  
  For any $c=\left(c_{0}, c_{1}, \ldots, c_{n-1}\right)\in \mathcal{C}$, 
 the code $\mathcal{C}$ is cyclic if the shifted codeword $c^\prime =\left(c_{n-1}, c_{0}, \ldots, c_{n-2}\right)$ also belongs to $\mathcal{C}$.

Let  $ \Omega_{n}=\{0,1, \ldots, n-1\}$, and let  $\gamma $  
be a primitive  $n$-th root of unity in some extension field of  $F_{4}$,  where $n$ is an odd integer. 
 The defining set $T$ of $\mathcal{C}=\langle g(x)\rangle$  
 is denoted as $T=\left\{i \in  \Omega_{n} \mid g\left(\gamma  ^{i}\right)=0\right\}$.

Let $i$ be an integer with $0\leq i < n$, the set $C_{i}=\{i, 4i,
4^{2}i, \cdots , 4^{k-1}i \}$ (mod  $n$) is called the
$4$-cyclotomic  coset
modulo $n$ that contains $i$, where $k$
is the smallest positive integer such that $4^{k}i$ $\equiv i$ (mod
$n$). For each $i \in \Omega_{n}$, a cyclotomic coset $C_{i}$ is called {\it skew symmetric} if $n-2i
$(mod $n$) belongs to $  C_{i}$, and is {\it skew asymmetric} otherwise.
Skew asymmetric cosets $C_{i}$ and $C_{n-2i}$ come in pair, we use
$(C_{i},C_{n-2i})$ to denote such a pair. Let $\mathcal {C}$ be a cyclic code with a defining set $T =
\bigcup_{i \in \Omega_{n}} C_{i}$. Denoting $T^{-2}=\{n-2x | x\in T \}$, then
we can deduce that the  defining set of $\mathcal {C}$$^{\bot _{h}}$
is $T^{\perp _{h}} =$$ \mathbf{b}{Z}_{n} $$\backslash T^{-2}$. If  $T \cap T^{-2}=\emptyset$, then  $\mathcal{C}^{\perp_{h}} \subseteq \mathcal{C}$, i.e., $g(x)\mid g^{\perp_h}(x)$.

 A code $\mathcal{C}$ is called quasi-cyclic of index $l$, or simply $l$-quasi-cyclic, 
   if shifting any codeword cyclically by $l$ positions results in another codeword 
   within $\mathcal{C}$. The length $n$ of such a quasi-cyclic code $\mathcal{C}$ 
   is a multiple of $l$, meaning $n = m \cdot l$ for some integer $m$. 
   Furthermore, by appropriately permuting the columns, the generator matrix of a quasi-cyclic 
   code can be arranged as an $m \times m$ block matrix, where each block is a circulant matrix.
It means that a 1-generator quasi-cyclic code can be transformed into an equivalent code with a generator matrix
$$ G = (G_{0}, G_{1}, G_{2}, \ldots,  G_{l-1}) $$
where  $A_{i}, i = 0,1,\ldots,l-1$ is   defined as $m\times m$ circulant matrix
$$
A=\left(\begin{array}{ccccc}
	g_{0} & g_{1} & g_{2} & \ldots & g_{m-1} \\
	g_{m-1} & g_{0} & g_{1} & \ldots & g_{m-2} \\
	g_{m-2} & g_{m-1} & g_{0}  & \ldots & g_{m-3} \\
	\vdots & \vdots & \vdots & \vdots & \vdots \\
	g_{1} & g_{2} & g_{3} & \ldots & g_{0}
\end{array}\right).$$

 With a suitable permutation of coordinates, the generator matrix of a 2-generator quasi-cyclic code 
 with  index $l$ can be transformed into the following form.
 $$G=\left(\begin{array}{ccccccc}
  G_{0,0} & G_{0,1}& G_{0,2}& \cdots & G_{0,l-1} \\
  G_{1,0} & G_{1,1}& G_{1,2}& \cdots & G_{1,l-1}\\
\end{array}\right)$$, where  $G_{i, j}$ is circulant matrices determined by polynomial  $a_{i, j}(x)$, 
where  $0 \leq i \leq 1$  and  $0\leq j \leq l-1.$

Let $g(x)=g_{0}+g_{1} x+g_{2} x+\cdots+g_{n-1} x^{n-1} \in \mathcal{R}$ and $[g(x)]=[g_{0},g_{1},g_{2},\cdots,g_{n-1}]$  represents vectors in  $F_{4}^{n}$  determined by the coefficient of $g(x)$ in an ascending order.

Let $g(x),h(x),\nu(x)$ be monic polynomials in $F_{4}[x]$ whose degree is less than $n$ 
and such that both $g(x),h(x)$ divide $x^{n}-1$. Any cyclic code of length $n$ and dimension
 $n-deg(g)$ can be generated by  $\langle g(x)\rangle$  .  
We consider the check polynomial $h(x)$ such that $g(x)\cdot h(x)=x^{n}-1$.Attached to a polymial $r(x)=r_{0}+r_{1}x+\dots+r_{m}x^{m}$ of degree 
$m<n$, we can define  $r^{[2]}(x)=r_{0}^{2}+r_{1}^{2}x+\dots+r_{m}^{2}x^{m}$.
Moreover, let  $ h(x)=(x^{n}-1)/g(x)$, then $g^{\perp _{h}}(x)=x^{deg(h(x))} h^{[2]}\left(\frac{1}{x}\right)$. 
It is well-known that $\langle g^{\perp _{h}}(x)\rangle$ generates the Hermitian dual code of cyclic code 
which is generated by $ \langle g(x)\rangle$.

\subsection{Binary Quantum codes}

A binary quantum error-correcting code $\mathcal{Q}$ of 
length $n$ is a $K$-dimensional subspace of $2^n$-dimensional Hilbert space $(\mathbf{C}^{2})^{\otimes n}$,
 where $\mathbf{C}$ denotes the complex field and $(\mathcal{C}^{2})^{\otimes n}$ is the $n$-fold tensor power 
 of $\mathcal{C}^2$. Such a binary quantum code $\mathcal{Q}$ is typically represented as $[[n,k,d]]_{2}$, where $k=log_{2}K$.

Hermitian construction is a well-known method for creating quantum codes, 
which connects quantum codes 
with classical self-orthogonal codes using the Hermitian inner product.

Lemma 2.1	(Corollary 19  \cite{Ketkar2006}, Hermitian construction) :
	If tere exists a  $[n, k, d]_{4}$ code  $\mathcal{C}$  such that $ C^{\perp_{h}} \subset \mathcal{C}$,
	 then there exists an  $[[n, 2 k-n, d]]_{2}$ quantum code that is pure to $d$, such that there are no 
	 vectors of weight $\le d-1$ in  $ C^{\perp_{h}} \setminus  \mathcal{C}$.

Quantum  codes can be derived from other quantum codes by the following propagation rules.

	Propagation2.2 (propagation rules\cite{Calderbank1998}): 
  If there exists an  $[[N, K, d]]_{2}$ quantum  codes,
	 then the following quantum  codes exist.\\
	(1)  $[[N, K-1, d]]_{2}$ for $K\ge 1$ ;\\
	(2)  $[[N+1, K, d]]_{2}$ for $K>0$ ;\\
	(3)  $[[N-1, K, d-1]]_{2}$ for $K>0$ \\
	(4)  $[[N-1, K+1, d-1]]_{2}$ for $N>2$ if the code is pure.

{\bf Notation 1.} In the sections that follow, within each generator matrix of linear codes, 
the symbols 2 and 3 are used to denote 
  $\omega$, and $\omega^2$, respectively.

\section{Construction of binary quantum codes}
In this section, we explore a class of 2-generator Hermitian dual-containing 
quasi-cyclic codes with index 2. Subsequently, 
by applying the Hermitian construction, numerous new quantum codes are developed.

	Definition 3.1. Let	$\nu(x)\in\mathcal{R}$ and $g_{1}(x)$, $g_{2}(x)$  be factors of  $x^{n}-1$. 
  Suppose that $A_{1}$, $A_{2}$, $A_{1}^{'}$, $A_{2}^{'}$   are generator matrices of cyclic codes  $\langle g_1(x)\rangle$, $\langle g_2(x)\rangle$, $\langle \nu(x)g_1(x)\rangle$ and $\langle \nu(x)g_2(x)\rangle$, respectively.
	Let	
	$$G=\left(\begin{array}{cc}A_{1}^{'}& A_1 \\A_2 & A_{2}^{'}\end{array}\right).$$ 
	Then, $\mathcal{C}_{4}(g_1,g_2,\nu)$  is a quasi-cyclic code with length $2n$ over  ${F}_{4}$   generated by matrix $G$. 

To build quantum codes with favorable parameters, 
it is necessary to find the dimension of the quasi-cyclic code described earlier.

%%%确定维数

Numerous quantum code constructions utilize self-orthogonal 
codes (also known as dual-containing codes) \cite{Calderbank1998,Ketkar2006}.
 Below, we present the key findings related to Hermitian self-orthogonality 
 with respect to the Hermitian inner product in the following lemma.
We  define the polynomial
$\bar{g}(x)= {\textstyle \sum_{i=0}^{n-1}g_{i}x^{n-i}} =g_{0}+g_{n-1} x+g_{n-2} x^{2}+ \cdots+g_{1} x^{n-1} $
 mod $(x^{n}-1)$.

   Lemma 3.2\label{Hermitian self-orthogonal} \cite{Bierbrauer2014TheSO},
  Let $\mathcal{C}$ be an quaternary linear code.  $\mathcal{C}$ is Hermitian self-orthogonal 
  if and only if $\langle c,c' \rangle_h=0$ for all codeword $c$ or
   $c'$ of $\mathcal{C}$ .

Lemma3.3 \label{exchange_law} \cite{galindo2018quasi}, 
Let  $f(x)$, $g(x)$  and  $h(x)$  be monic polynomials in  $\mathcal{R}$. Then the vectors corresponding to the coefficients  of the  polynomials have the following equality of the Hermitian inner product:
$$\langle[f(x) g(x)],[h(x)]\rangle_{h}=\left\langle[g(x)],
\left[\overline{f}(x) h(x)\right]\right\rangle_{h}.$$

Let $$G^{\perp}=\left(\begin{array}{cc}
    A^{'\perp}_{1}& A^{\perp}_1 \\
    A^{\perp}_2 & A^{'\perp}_{2}
  \end{array}\right),$$
where	$A^{\perp}_{1}$, $A^{\perp}_{2}$, 	$A^{'\perp}_{1}$ and $A^{'\perp}_{2}$  are generator matrices of cyclic codes  generated by $[g^{\perp_h}_1(x)]$,  $[g^{\perp_h}_2(x)]$, $[-\bar{\nu}(x)g^{\perp_h}_1(x)]$ and $[-\bar{\nu}(x)g^{\perp_h}_2(x)]$, respectively.	
Then, $\mathcal{C}^{\perp}=[2n,deg(g_1(x))+deg(g_2(x))]$  is defined as a quasi-cyclic code over  ${F}_{4}$  of length  $2n$ with its generator matrix $G^{\perp}$.

Lemma3.4 \label{innerproductzero}\cite{Guan2022QC}, 
	Let  $\mathcal{C}_{g_{1}}$ and $\mathcal{C}_{g_{2}}$ are linear codes of length $n$ with generator  polynomial $g_{1}(x)$ and $g_{2}(x)$, respectively. 
	The   $\langle[a(x)g_{1}(x)],[b(x)g_{2}(x)]\rangle_{h}=0$  for any polynomials $a(x)$ and $b(x)$ in $\mathcal{R}$,
	 if and only if $g_{2}^{\perp_h} (x)\mid g_{1}(x)$ and $g_{1}^{\perp_h} (x)\mid g_{2}(x)$.

	From Lemma3.4, we can give the dual code of the quasi-cyclic
	 code in Definition 1 as above.

%%给出两个码互为对偶码的条件
Proposition 3.5\cite{Guan2022QC}\label{1dual_code} 
		Let $\mathcal{C}=\left[2 n,2n-deg\left(g_{1}(x)\right)
		-deg\left(g_{2}(x)\right)\right]_{4}$ is defined in Definition 1. 
		 $\mathcal{C}^{\perp_h}=\left[2n,deg\left(g_{1}(x)\right)+
		 deg\left(g_{2}(x)\right)\right]_{4}$ are given as above. 
		 Then  $\mathcal{C}^{\perp_h}$ is the Hermitian dual code of $\mathcal{C}$.

%%给出对偶包含条件的判定

Lemma 3.6 \cite{galindo2018quasi,Guan2022QC}\label{dual-containing} 
	Let $\mathcal{C}$ and $\mathcal{C}^{\perp_h}$ be defined as 
Proposition \ref{1dual_code}.	If $g_{1}(x)$, $g_{1}^{\perp_h}(x)$, $g_{2}(x)$ , $g_{2}^{\perp_h}(x)$, 
and $\nu(x)$ which defined in Definition 3.1 satisfies  Lemma \ref{innerproductzero},
 then $\mathcal{C}$ is Hermitian dual-containing or $\mathcal{C}^{\perp_h}$ is Hermitian self-otrhogonal.

Theorem 3.7 \label{method} Let  $\mathcal{C}(g_1,g_2,\nu)$ 
	 be a quasi-cyclic code with dimension of $ 2 n-deg\left(g_{1}(x)\right)-deg
   \left(g_{2}(x)\right)$ as described in Definition  3.1 satisfy Lemma 
	 \ref{dual-containing}. Then there exists a pure binary quantum code
	  $\left[\left[2n, 2n-2deg\left(g_{1}(x)\right)-2deg\left(g_{2}(x)\right),
	   d\right]\right]$, 
	  where  $d=\min \left\{wt(\vec{c}) \mid \vec{c} \in \mathcal{C}(g_1,g_2,\nu)\right\}$.

Proof:
By Definition  3.1 and Lemma \ref{dual-containing}, we can construct a dual-containing
 quasi-cyclic code $\mathcal{C}=[2n,k,d]_{4}$ with parameter 
 $[2n,2n-deg\left(g_{1}(x)\right)-deg\left(g_{2}(x)\right),d]_{4}$.
 According to Lemma   2.1, then there exsits a  
 $\left[\left[2n, 2n-2deg  \left(g_{1}(x)\right)-2deg\left(g_{2}(x)\right), 
 d\right]\right]_{2}$ 
 quantum code.

Let  two cyclic codes $\mathcal{C}_1$,  $\mathcal{C}_2$  be generated by  $\left\langle g_{1}(x)\right\rangle$, $\left\langle g_{2}(x)\right\rangle$ 
with defining sets  $T_{1}$, $T_{2}$ , respectively.
If $T_{1} \cap T^{-2}_{1}=\emptyset$ and $T_{2} \cap T^{-2}_{2}=\emptyset$, then  $g_{1}(x) 
\mid g_{1}^{\perp_h}(x)$  and $g_{2}(x) \mid g_{2}^{\perp_h}(x)$. Given  the appropriate $\nu$, one can construct 
a code  $\mathcal{C}_{4}(g_1,g_2,\nu)$$=\left[2 n, 2 n-deg
\left(g_{1}(x)\right)-deg\left(g_{2}(x)\right)\right]$ and the dual $\mathcal{C}^{\perp_h}$$=\left[2 n, 
deg \left(g_{1}(x)\right)+deg\left(g_{2}(x)\right)\right]$.
$\mathcal{C}_{4}(g_1,g_2,\nu)$ is dual-containing and $\mathcal{C}^{\perp_h}$ is self-orthogonal.
By  an exhaustive search of calculation with Magma \cite{bosma1997magma}, we can construct 
 good dual-containing codes $\mathcal{C}_{4}(g_1,g_2,\nu)$ with improved minimal distance $d$.
 Because the Hermitian construction for quantum codes 
from Hermitian dual-containing codes keeps the minimal distance of the code and its dual, we can construct 
many quantum codes with paraments
$\left[\left[2 n, 2 n-2{deg}
\left(g_{1}(x)\right)-2{deg}\left(g_{2}(x)\right),d\right]\right]_{2}$ which  
improve the currently best-known  ones in \cite{Grassl:codetables}.

We express coefficient polynomials in ascending order and use indexes of elements to express the 
same number of consecutive elements. For example, 
$1+\omega^{2}x+\omega x^4+x^5$ can be presented as $130^221$.

%%给出两个码互为对偶码的条件
Proposition 3.8 \label{1dual_code} 
  There exist binary quantum codes with parameters:

  [[42,10,9]],[[42,9,9]],[[43,10,9]],[[44,10,9]],[[45,10,9]],[[74,36,9]],
  [[74,35,9]],[[74,34,9]],[[74,33,9]],[[75,36,9]],[[75,35,9]],[[75,34,9]],
  [[75,33,9]],[[76,36,9]],[[76,35,9]],[[76,34,9]],[[76,33,9]],[[77,36,9]],
  [[77,35,9]],[[77,34,9]], [[78,40,9]],[[78,39,9]],[[79,40,9]],[[79,39,9]],
  [[80,40,9]],[[80,39,9]],[[77,41,8]].

Proof:
  
  1. Let  $n=21$. Examine the 4-cyclotomic cosets modulo 21. 
  For  $T_{1}=C_{3}$ and  $T_{2}=C_{1}\cup C_{2}\cup C_{3}\cup C_{5}\cup C_{14}$  as the defining sets of cyclic codes 
   $\left\langle g_{1}(x)\right\rangle$  and  $\left\langle g_{2}(x)\right\rangle$. 
   Then   $g_{1}(x)=x^{3}+x + 1$, which is presented as $g_1(x)=1101$.
   $g_{2}(x)=x^{13} + wx^{12} + w^2x^9 + wx^8 + w^2x^6 + x^5 + w^2x^4 
   + x^3 + w^2x^2 + 1$,
   which is presented as $g_2(x)=3^{2}1^{2}(31)^{2}12(21)^{2}$.
   
   Using Magma for the calculation, we select $\nu(x)=w^2x^{20} + x^{19} + x^{18} + w^2x^{17} + w^2x^{16} + x^{15} + w^2x^{14} 
 + w^2x^{13} + x^{12} + wx^{11} + x^{10} + w^2x^9 + x^8 + 
 w^2x^6 + wx^5 + x^4 + wx^2 + wx + 1$, which is presented as $\nu(x)=1020231213^{2}23^{2}0332^{3}$.  
 Based on Proposition \ref{1dual_code}, we have constructed a quaternary
  Hermitian dual-containing quasi-cyclic code with parameters  $[42, 26, 9]_{4}$ 
  with generator matrix of $G_{42}$. Its weight distribution is given by  
 $w(z)=1+3486 z^{9}+22176 z^{10}+181566 z^{11}+\cdots+ 356314153608 z^{41}+
  25503008994 z^{42}$.

A Hermitian self-orthogonal quasi-cyclic code 
with parameters $[42,16,12]_4$ serves as its dual code. 
Using the Hermitian construction, a new binary quantum code with parameters 
$[[42, 10, 9]]_2$ has been developed. According to Grassl's code tables 
\cite{Grassl:codetables}, the best-known binary quantum code with these 
parameters previously had a minimum distance of $8$. 
Therefore, the current result improves the known minimum distance to $9$.

Proposition 2.2 gives us 4 codes, with respective parameters
$[[42, 9, 9]]_{2}$ ,$[[43, 10, 9]]_{2}$, $[[44,10,9]]_{2}$ and $[[45,10, 9]]_{2}$. They are better than the codes with parameters
$[[42, 9, 8]]_{2}$ ,$[[43, 10, 8]]_{2}$, $[[44,10,8]]_{2}$ and $[[45,10, 8]]_{2}$  which held the previous record.

2.  $[[74, 36, 9]]_{2}$: Examine the $4$-cyclotomic cosets modulo $37$. For $T_{1}=C_{1};
 $ and  $T_{2}=C_{0}$  as the defining sets of cyclic codes  
 $\left\langle g_{1}(x)\right\rangle$  and  $\left\langle g_{2}(x)\right\rangle$. 
 Then   $g_{1}(x)=x^{18} + w^2x^{17} + x^{16} + x^{15} + w^2x^{14} + w^2x^{13} + wx^{12} + x^{11}
  + w^2x^{10} + x^9 + w^2x^8 + x^7 + wx^6 + 
 w^2x^5 + w^2x^4 + x^3 + x^2 + w^2x + 1$, which is presented as 
 $g_1(x)= 131^{2}3^{2}2(13)^{2}123^{2}1^{2}31$,
 $g_{2}(x)=x + 1$, which is presented as $g_2(x)=1^{2}$.
  Using Magma for the calculation, we select 
  $\nu(x)=x^{36} + wx^{34} + wx^{33} + x^{32} + wx^{31} + x^{30} + w^2x^{29} + wx^{28} + 
  x^{27} + w^2x^{26}
   + x^{25} + x^{24} + x^{23} + w^2x^{22} +
  w^2x^{21} + wx^{19} + w^2x^{18} + wx^{17 }+ wx^{16} + x^{15} + w^2x^{14} + w^2x^{13} 
  + wx^{12} + wx^{11} + wx^{10 }
  + x^9 + w^2x^8 +
  x^7 + w^2x^6 + wx^5 + x^4 + x^3 + x^2 + 1$, which can be presented as
  $\nu(x)=101^{3}2(31)^{2}2^{3}3^{2}12^{2}3203^{2}1^{3}3123(12)^{2}201$.

  According to  Proposition \ref{1dual_code}, we can obtain a Hermitian dual-containing code $[74,55,9]_{4}$ with generator matrix of $G_{74}$, whose weight distribution is
   $w(z)=1+ 9213z^{9}+153180 z^{10}+2700408 z^{11}+\cdots+737620560929545709890785 z^{74}$.
 
 Using the Hermitian construction, a new binary quantum code with 
 parameters $[[74, 36, 9]]_2$ has been developed. According to Grassl's 
 code tables \cite{Grassl:codetables}, the best previously known binary 
 quantum code with these parameters was $[[74, 36, 8]]_2$. 
 Therefore, the current record for the minimum distance has been improved to 9.
  
  Proposition 2.2 gives us 14 codes, with respective parameters

  $[[74, 35, 9]]_{2}$, $[[74, 34, 9]]_{2}$, $[[74, 33, 9]]_{2}$, $[[75, 36, 9]]_{2}$, $[[75, 35, 9]]_{2}$,
   $[[75, 34, 9]]_{2}$, $[[75, 33, 9]]_{2}$, 
   $[[76, 36, 9]]_{2}$, $[[76, 35, 9]]_{2}$, $[[76, 34, 9]]_{2}$,
    $[[76, 33, 9]]_{2}$, 
   $[[77, 36, 9]]_{2}$, $[[77, 35, 9]]_{2}$ and $[[77, 34, 9]]_{2}$. They are better than the codes with parameters
  $[[74, 35, 8]]_{2}$, $[[74, 34, 8]]_{2}$, $[[74, 33, 8]]_{2}$, $[[75, 36, 8]]_{2}$, $[[75, 35, 8]]_{2}$,
   $[[75, 34, 8]]_{2}$, $[[75, 33, 8]]_{2}$, $[[76, 36, 8]]_{2}$, $[[76, 35, 8]]_{2}$, 
   $[[76, 34, 8]]_{2}$, $[[76, 33, 8]]_{2}$, $[[77, 36, 8]]_{2}$, $[[77, 35, 8]]_{2}$ and $[[77, 34, 8]]_{2}$ 
    which held the previous record.

 3. $[[ 78,40,9]]_{2}$: Consider the $4$-cyclotomic cosets modulo $39$. Select  $T_{1}=C_{1}\cup C_{2}\cup C_{6};
 $ and  $T_{2}=C_{26}$  as the defining sets of cyclic codes  $\left\langle g_{1}(x)\right\rangle$  and  $\left\langle g_{2}(x)\right\rangle$. 
 
  Then  $g_{1}(x)=x^18 + wx^{17} + wx^{16} + x^{14} + x^{13} + w^2x^{11} + 
 w^2x^9 + x^7 + w^2x^5 + wx^4 + x^2 + w^2x + 1,$, which can be presented as 
 $g_1(x)=13102301(03)^{2}01^{2}02^{2}1$.  $g_{2}(x)=x + w^2$, which can be presented as $g_2(x)=31$.
 and $\nu(x)=w^2x^{38} + wx^{37} + w^2x^{36} + wx^{34} + x^{32} + 
 w^2x^{31} + w^2x^{30} + wx^{29} + x^{28} + w^2x^{27} + w^2x^{26} + w^2x^{24} +
 w^2x^{23} + w^2x^{22} + x^{21} + wx^{20} + w^2x^{19} + w^2x^{18} + 
 w^2x^{16} + wx^{15} + wx^{14} + wx^{13} + wx^{12} + w^2x^{11} + 
 w^2x^9 + wx^7 + x^6 + w^2x^5 + x^3 + wx + w$, which can be presented as
 $\nu(x)=22010312(03)^{2}$ $2^{4}303^{2}213^{3}03^{2}123^{2}1020323$. 
 Based on Proposition \ref{1dual_code}, we have constructed a quaternary
  Hermitian dual-containing quasi-cyclic code with parameters   $[78, 59, 9]_{4}$  
  with generator matrix of $G_{78}$.
   Its weight distribution is  $w(z)=1+13806 z^{9}+258219 z^{10}+5034744 z^{11}+83901051 z^{12}+\cdots+ 
 59747265434538325974318951 z^{78}$.	
 
Proposition 2.2 gives us 6 codes, with respective parameters
$[[78, 39, 9]]_{2}$, $[[79, 40, 9]]_{2}$, $[[79, 39, 9]]_{2}$, $[[80, 40, 9]]_{2}$, $[[80, 39, 9]]_{2}$  
and $[[77,41,8]]_{2}$. 
They are better than the codes with parameters
$[[78, 39, 8]]_{2}$, $[[79, 40, 8]]_{2}$, $[[79, 39, 8]]_{2}$, $[[80, 40, 8]]_{2}$, $[[80, 39, 8]]_{2}$  
and $[[77,41,7]]_{2}$  
which held the previous record.

 %%%%%%%%%%%%%%%%%%%%%%%%%%%%%%%%%%%%%%%%%%%%--------四元码-------%%%%%%%%%%%%%%%%%%%%%%%%%%%%%%%%%%%%%%%%%%%%%%%%%%%%
 The constructions dual-containing quasi-cyclic codes 
 in this paper are listed in Table \ref{tab: quasi-cyclic}.
 
  \begin{table}[h]
   \caption{Dual-containing quasi-cyclic codes with better paraments}
   \label{tab: quasi-cyclic}
   \begin{center}

     \begin{tabular}{ccc}
       \hline
          $\mathcal{C}_4(g_1,g_2,\nu)$      & $g_1(x),\quad g_2(x),\quad\nu(x)$\\ \hline

   \hline
   $[42,26,9]_{4}$          & $1101$, \\
                            &   $3^{2}1^{2}(31)^{2}12(21)^{2}$, \\
                            &   $ 1020231213^{2}23^{2}0332^{3}$        \\
  \hline
  $[74,55,9]_{4}$          & $ 131^{2}3^{2}2(13)^{2}123^{2}1^{2}31$, \\
  &$ 1^{2}$, \\
  &$101^{3}2(31)^{2}2^{3}3^{2}12^{2}3203^{2}1^{3}3123(12)^{2}201$  \\
 \hline
 
   $[78,59,9]_{4}$          & $31$, \\
                            & $13102301(03)^{2}01^{2}02^{2}1$, \\
                            & $ 22010312(03)^{2}2^{4}303^{2}213^{3}03^{2}123^{2}1020323$  \\

   \hline
     \end{tabular}\end{center}
   \end{table}

   From these dual-containing quasi-cyclic codes,
   We have found 3 new quantum codes derived from quasi-cyclic codes, together with 24 derived codes,
   which improve  the lower bounds on the minimum distance in Grassl's table
  \cite{Grassl:codetables}.  All the new  improved quantum codes in this paper are shown in Table \ref{tab: binary}, 
  whose parameters  improve the lower bounds on the minimum distance in Grassl's table \cite{Grassl:codetables}.

   \begin{table}[h]
   \caption{New binary quantum codes}
   \label{tab: binary}
   \begin{center}

     \begin{tabular}{ccc}
       \hline
       NO. &    Our Codes      & Codes in Grassl's table \cite{Grassl:codetables} \\ \hline
 
     1  & $[[42,10,9]]_{2}$  &     $  [[42,10,8]]_{2}$      \\
       2  & $[[42,9,9]]_{2}$  &      $  [[42,9,8]]_{2}$      \\ 
       3  & $[[43,10,9]]_{2}$  &     $  [[43,10,8]]_{2}$      \\      
       4  & $[[44,10,9]]_{2}$  &     $  [[44,10,8]]_{2}$      \\ 
     5  & $[[45,10,9]]_{2}$  &     $  [[45,10,8]]_{2}$      \\     
     
    6 & $[[74,36,9]]_{2}$ &      $[[74,36,8]]_{2}$       \\ 
    7  & $[[74,35,9]]_{2}$ &      $[[74,35,8]]_{2}$       \\ 
    8  & $[[74,34,9]]_{2}$ &      $[[74,34,8]]_{2}$       \\
    9  & $[[74,33,9]]_{2}$ &      $[[74,33,8]]_{2}$       \\
    10  & $[[75,36,9]]_{2}$ &      $[[75,36,8]]_{2}$       \\ 
    11  & $[[75,35,9]]_{2}$ &      $[[75,35,8]]_{2}$       \\ 
    12  & $[[75,34,9]]_{2}$ &      $[[75,34,8]]_{2}$       \\
    13  & $[[75,33,9]]_{2}$ &      $[[75,33,8]]_{2}$       \\
      14  & $[[76,36,9]]_{2}$ &      $[[76,36,8]]_{2}$       \\
      15  & $[[76,35,9]]_{2}$ &      $[[76,35,8]]_{2}$       \\ 
      16  & $[[76,34,9]]_{2}$ &      $[[76,34,8]]_{2}$       \\
      17  & $[[76,33,9]]_{2}$ &      $[[76,33,8]]_{2}$       \\ 
    18  & $[[77,36,9]]_{2}$ &      $[[77,36,8]]_{2}$       \\
    19  & $[[77,35,9]]_{2}$ &      $[[77,35,8]]_{2}$       \\ 
    20  & $[[77,34,9]]_{2}$ &      $[[77,34,8]]_{2}$       \\

    21  & $[[78,40,9]]_{2}$ &    $[[78,40,8]]_{2}$       \\
    22  & $[[78,39,9]]_{2}$ &    $[[78,39,8]]_{2}$       \\
    23  & $[[79,40,9]]_{2}$ &    $[[79,40,8]]_{2}$       \\
    24  & $[[79,39,9]]_{2}$ &    $[[79,39,8]]_{2}$       \\
    25  & $[[80,40,9]]_{2}$ &    $[[80,40,8]]_{2}$       \\
    26  & $[[80,39,9]]_{2}$ &    $[[80,39,8]]_{2}$       \\
    27  & $[[77,41,8]]_{2}$ &    $[[77,41,7]]_{2}$       \\

   \hline
     \end{tabular}\end{center}
   \end{table}
   
 However, even supercomputers find it difficult to determine the 
 specific parameters of quasi-cyclic codes
  for larger code lengths and dimensions.

\section{SUMMARY}

This paper investigates a class of quaternary 2-generator quasi-cyclic codes with index 2. 
One of the challenges for these 2-generator quasi-cyclic codes is 
how to determine the minimum distance of the code.
  By an exhaustive search using MAGMA, We determine 
 the dual-containing  quasi-cyclic codes under the Hermitian inner product with good minimum distance. 
  We have found 3 new quantum codes $[[42,10,9]]$, $[[74,,36,9]]$ and 
  $[[78,40,9]]$  derived from the quasi-cyclic codes, 
 together with 24 new derived codes   
 $[[42,9,9]], [[43,10,9]]$, $[[44,10,9]]$,$[[45,10,9]]$, $[[74,35,9]],[[74,34,9]],[[74,33,9]]$,
 $[[75,36,9]],$  $[[75,35,9]],$ $[[75,34,9]],[[75,33,9]]$,$[[76,36,9]], [[76,35,9]],[[76,34,9]],$ $[[76,33,9]],
 [[77,36,9]],  [[77,35,9]]$,$[[77,34,9]],[[78,39,9]],$ $[[79,40,9]], [[79,39,9]],[[80,40,9]],
 [[80,39,9]]$,$[[77,41,8]]$ ,
 which improve the
 best-known lower bounds on minimum distance in Grassl's code tables \cite{Grassl:codetables}.
 
  Through exhaustive search in MAGMA, more quantum codes can be  found  
 which     their minimum distances beat the
 minimum distances of the previously known quantum  codes  in Grassl's code tables. 
 There are some interesting things. 
 Why the codes with some elections of $f_{1},f_{2},\nu(x)$ are better than others? 
 We  hope to find the reasons and regular pattern in our next future works.
 For given $f_{1},f_{2}$, what properties do $\nu(x)$  need to the 
 constructed code satisfies self orthogonality  (or dual-containing conditions)?
 What is the relationship between the elections of $f_{1},f_{2},\nu(x)$
  and the dimension of the quasi-cyclic code determined by the Gr$\ddot{o}$bner bases?

%\section*{Acknowledgment}

%\section*{References}

\vspace{12pt}
\color{red}

\end{document}